\documentclass[aps,amsmath,amssymb,reprint,superscriptaddress]{revtex4-2}

\usepackage{graphicx}
\usepackage{subfigure}
\usepackage{bm}
\usepackage{lineno}
\usepackage{color}
\usepackage{verbatim}
\usepackage{soul}

\modulolinenumbers[1]
\begin{document}

\preprint{APS/123-QED}

\title{\boldmath The Feasibility Study of the GeV-Energy Muon Source Based on HIAF}

\author{Yu Xu}
\affiliation{Advanced Energy Science and Technology Guangdong Laboratory, Huizhou 516000, China}
\affiliation{Institute of Modern Physics, CAS, Lanzhou 730000, China}

\author{Xueheng Zhang}
\affiliation{Advanced Energy Science and Technology Guangdong Laboratory, Huizhou 516000, China}	
\affiliation{Institute of Modern Physics, CAS, Lanzhou 730000, China}
\affiliation{School of Nuclear Science and Technology, University of Chinese Academy of Sciences, Beijing 100049, China}

\author{Yuhong Yu}
\affiliation{Advanced Energy Science and Technology Guangdong Laboratory, Huizhou 516000, China}	
\affiliation{Institute of Modern Physics, CAS, Lanzhou 730000, China}
\affiliation{School of Nuclear Science and Technology, University of Chinese Academy of Sciences, Beijing 100049, China}

\author{Pei Yu}
\affiliation{Advanced Energy Science and Technology Guangdong Laboratory, Huizhou 516000, China}
\affiliation{Institute of Modern Physics, CAS, Lanzhou 730000, China}

\author{Li Deng}
\affiliation{Advanced Energy Science and Technology Guangdong Laboratory, Huizhou 516000, China}
\affiliation{Institute of Modern Physics, CAS, Lanzhou 730000, China}

\author{Jiajia Zhai}
\affiliation{Advanced Energy Science and Technology Guangdong Laboratory, Huizhou 516000, China}
\affiliation{Institute of Modern Physics, CAS, Lanzhou 730000, China}

\author{Liangwen Chen}
\email{chenlw@impcas.ac.cn}
\affiliation{Advanced Energy Science and Technology Guangdong Laboratory, Huizhou 516000, China}	
\affiliation{Institute of Modern Physics, CAS, Lanzhou 730000, China}
\affiliation{School of Nuclear Science and Technology, University of Chinese Academy of Sciences, Beijing 100049, China}

\author{He Zhao}
\affiliation{Advanced Energy Science and Technology Guangdong Laboratory, Huizhou 516000, China}	
\affiliation{Institute of Modern Physics, CAS, Lanzhou 730000, China}
\affiliation{School of Nuclear Science and Technology, University of Chinese Academy of Sciences, Beijing 100049, China}

\author{Lina Sheng}
\affiliation{Advanced Energy Science and Technology Guangdong Laboratory, Huizhou 516000, China}	
\affiliation{Institute of Modern Physics, CAS, Lanzhou 730000, China}
\affiliation{School of Nuclear Science and Technology, University of Chinese Academy of Sciences, Beijing 100049, China}

\author{Guodong Shen}
\affiliation{Advanced Energy Science and Technology Guangdong Laboratory, Huizhou 516000, China}	
\affiliation{Institute of Modern Physics, CAS, Lanzhou 730000, China}
\affiliation{School of Nuclear Science and Technology, University of Chinese Academy of Sciences, Beijing 100049, China}

\author{Ziwen Pan}
\affiliation{State Key Laboratory of Particle Detection and Electronics, University of Science and Technology of China, Hefei 230026, China}

\author{Qite Li}
\affiliation{State Key Laboratory of Nuclear Physics and Technology,
	School of Physics, Peking University, Beijing, 100871, China}

\author{Chen Zhou}
\affiliation{State Key Laboratory of Nuclear Physics and Technology,
	School of Physics, Peking University, Beijing, 100871, China}

\author{Qiang Li}
\affiliation{State Key Laboratory of Nuclear Physics and Technology,
	School of Physics, Peking University, Beijing, 100871, China}

\author{Lei Yang}
\affiliation{Advanced Energy Science and Technology Guangdong Laboratory, Huizhou 516000, China}	
\affiliation{Institute of Modern Physics, CAS, Lanzhou 730000, China}
\affiliation{School of Nuclear Science and Technology, University of Chinese Academy of Sciences, Beijing 100049, China}

\author{Zhiyu Sun}
\email{sunzhy@impcas.ac.cn}
\affiliation{Advanced Energy Science and Technology Guangdong Laboratory, Huizhou 516000, China}	
\affiliation{Institute of Modern Physics, CAS, Lanzhou 730000, China}
\affiliation{School of Nuclear Science and Technology, University of Chinese Academy of Sciences, Beijing 100049, China}

\date{\today}

\begin{abstract}

Generating a mono-energetic, high-energy muon beam using accelerator facilities can be very attractive for many purposes, for example, improving muon tomography currently limited by the low flux and wide energy spread of cosmic ray muons, and searching for muon related new physics beyond the Standard Model.
One potential accelerator facility is the High Intensity Heavy-Ion Accelerator Facility (HIAF), which is currently under construction in Huizhou City, China.
Considering the projectile energy and beamline length, a high-intensity and GeV-energy muon flux could be produced and delivered by the High Energy Fragment Separator beamline of the HIAF facility. In this paper, the flux intensity and purity of muon beam based on HIAF are discussed in detail. 
For the $\mu^+$ beam, the highest muon yield reaches $8.2 \times 10^6 ~ \mu$/s with the purity of approximately $2\%$ at a momentum of 3.5 GeV/c; meanwhile, for the $\mu^-$ beam, the maximum muon yield is 4.2 $\times 10^6 ~ \mu$/s with the purity of around $20\%$ at a momentum of 1.5 GeV/c. The results also indicate that, for muon beams with an energy of several GeV, by applying a suitable purification strategy, we can get a muon beam with a purity of 100\% and an intensity of the order of $10^5 ~ \mu$/s.

\end{abstract}

\maketitle

\section{\label{sec:Introduction}Introduction}

Muon tomography is a very unique tool for the inspection and imaging of large and dense objects, where other techniques often fail. Muons do not undergo strong interactions and the corresponding energy loss is much less than that of electron and proton when passing through matter \cite{Zhong:2016pbb}, rendering muons with great penetrating ability. For this reason, muon tomography technique has drawn great attention over the past several decades.

Cosmic ray muons are the conventional muon sources applied in muon tomography. In 2003, cosmic ray muons were successfully demonstrated to image medium-to-large and dense objects \cite{Borozdin2003}. Since then, muon tomography has been regarded as a promising tool to reconstruct a three dimensional image of the density of extended volumes. In addition, the relevant technologies of muon tomography have undergone rapid developments. Currently, muon tomography has been widely used for diverse applications in various scientific fields, for example, the geological imaging of volcanos\cite{tanaka2019,Tioukov2019}, non-destructive exploration of historical sites\cite{AlMoussawi:2023uji}, monitoring nuclear waste containers\cite{Perry:2013bja}, commercial and security inspection of dense radioactive materials in cargo containers\cite{georgadze2024,Baesso:2014xna}.  
Meanwhile, an increasing number of applications of muon tomography are emerging.
However, the application of muon tomography is still hampered by the low flux and wide energy spread of cosmic ray sources. Since the cosmic ray muon flux is approximately 1 muon $\text{cm}^{-2}\text{min}^{-1}$ with the energy range from sub-GeV to several hundred TeV at the Earth's surface~\cite{Durham:2017zfw}, most muon tomography applications require a considerable amount of time and the relevant resolution is limited, making it rather  impractical. 

Furthermore, accelerator facilities also can produce the GeV-energy muon beams, which have the potential for application in muon tomography. Despite this potential, there are no practical implementations of muon tomography utilizing accelerator-generated muon beams by now. The muon g-2 experiment at the Fermilab muon campus can provide a muon beam with a momentum of 3.1 GeV/c \cite{Muong-2:2021ojo}. However, this beam is not suitable for muon tomography purposes due to its pulsed nature. Similarly, the M2 Beamline at CERN can produce muon beams with energies reaching up to several hundred GeV \cite{NA64:2024klw}. Such high energies render the beam unsuitable for muon tomography applications, due to its excessively high penetrating capability.
Furthermore, muon beams with energies in the range of several GeV, which are generated through the decay of pions in the decay pipes at neutrino facilities, are also viable candidates for muon tomography \cite{Suerfu_2016}. Ref. \cite{Suerfu_2016} has explored the possibility of achieving high-resolution muon tomography using multi-GeV muon beams, highlighting the potential of these beams for this specific application.

In addition to the cosmic ray muons and accelerator muon sources, extensive efforts have been made to develop various high-energy, high-intensity muon sources for muon tomography. 
In Ref. \cite{NAGAMINE2004}, a muon source for advanced muon radiography was proposed. It is based on a compact acceleration system that integrates electron accelerator, production, transportation, cooling and rapid re-acceleration of muons. In that proposal, an accelerated muon source can be provided with mono-energetic, high-energy, high-intensity and pencil-like beam profiles. 
Furthermore, Ref. \cite{Liu:2025ejy} recently proposed a novel muon source concept employing a high-repetition-rate pulsed electron beam, such as that available at the SHINE facility. This electron beam-induced muon generation scheme shows particular promise for applications in muon tomography.
In recent years, laser-driven muon sources have attracted attention due to their promising potential for producing bright and compact muon beams\cite{dimu09,Calvin}. Ref.\cite{dimu09} proposed a table-top configuration to generate high-energy muons with a PW laser facility. Numerically, Ref. \cite{Calvin} has shown that with a high-power laser facility, a high-flux of muons that could meet the requirements for meaningful tomography applications can be generated.
Recently, muons have been experimentally obtained using ultra-short, high-intensity lasers \cite{zhang2024,terzani2024}, which will provide an exciting future for muon tomography with high-energy, high-intensity muon beams. Meanwhile, it has been suggested that the accelerated muon beam could be high-intensity, monochromatic and high-energy. This kind of accelerated muon beam is critical to achieve imaging with high resolution in less time\cite{Otani:21}. Acceleration of muons by a radio-frequency accelerator has also been demonstrated for the first time\cite{MUA}, which would pave the way for breakthroughs in muon tomography. 
All these efforts of high-energy and high-intensity muons would unlock the potential of muon imaging, and thereby open a new era of precision muon tomography. 

Moreover, experiments with GeV-energy muons are expected to explore various new physics beyond the Standard Model. For instance, muons with GeV energies can undergo large-angle scattering with potential muon-philic dark matter\cite{PhysRevD.110.016017}, making them a powerful tool for investigating possible interactions between dark matter and muons. They can also probe new gauge bosons associated with charged lepton flavor violation with muon-electron scattering\cite{Gao:2024xvf}. Moreover, precise measurements of GeV muon-electron scattering offer a unique opportunity to test quantum entanglement\cite{Gao:2024leu}.

The High Intensity heavy-ion Accelerator Facility (HIAF)\cite{HIAF}, currently under construction at the Institute of Modern Physics (IMP) in China, can provide ion beams (ranging from proton to uranium)  with the energies on the order of GeV/u and intensities as large as $10^{11} \sim 10^{12}$ particles per pulse (ppp). The energy of the primary beam is large enough for the production of pions. In addition, the High energy FRagment Separator (HFRS)\cite{Sheng:2023ojn,Wu:2024bko}, a major sub-system of the accelerator complex of HIAF, is designed to produce, separate, identify, and transport unstable nuclides of interest for various investigations. 
The HFRS is 192 $\mathrm{m}$ long and can operate with the magnetic rigidity ($B\rho$) up to 25 Tm, enabling the transportation of pions and muons with momenta up to $7.5$ $\mathrm{GeV/c}$, in accordance with the formula $B\rho[\mathrm{Tm}] = p[\mathrm{MeV}/\mathrm{c}]/300\,q$, where $p$ represents the momentum of the particle, and $q$ is the charge of the particle. Given that the decay length of pion with momentum $p_{\pi}$ is expressed as $L_{\pi}[\mathrm{cm}] = 5.593 \,\times \, p_{\pi} [\mathrm{MeV/c}]$ \cite{Nagamine_2003}, the decay length of a 3 $\mathrm{GeV/c}$ pion is about 170 $\mathrm{m}$. Therefore, the HFRS beamline is long enough to allow the decay of pions in the GeV energy range, making it possible to produce muons with energies of several GeV. 
However, the HFRS is a meticulously-designed separator tailored specifically for radioactive beam experiments. Consequently, compared with dedicated muon beamlines, it has relatively limited acceptance in angular and momentum.
In light of these characteristics, the present paper aims to conduct a comprehensive investigation into the muon yield and purity of the HFRS. The objective of this study is to explore the feasibility of utilizing the HFRS for muon beam generation.

\section{Brief overview of HIAF facility}

\begin{table}[h]
	\centering
	\caption{Typical beam parameters for the BRing\cite{HIAF}.}
	\label{tab:beam_param}
	\begin{tabular}{lcc}
		\hline
		\hline
		Ion species & Energy/(GeV/u) & Intensity/(ppp) \\
		\hline
		$^{238}\rm{U}^{35+}$   & 0.835   & $1.0\times 10^{11}$ \\
		$^{209}\rm{Bi}^{31+}$  & 0.85    & $1.2\times 10^{11}$ \\
		$^{78}\rm{Kr}^{19+}$   & 1.7    & $3.0\times 10^{11}$ \\
		$^{18}\rm{O}^{6+}$     & 2.6    & $6.0\times 10^{11}$ \\
		$\rm{proton}$          & 9.3    & $2.0\times 10^{12}$ \\
		\hline
		\hline
	\end{tabular}
\end{table}

HIAF, a large-scale scientific facility, mainly consists of the ion source, the superconducting ion linac accelerator (iLinac), the synchrotron booster ring (BRing) \cite{Zhang:2024aye,Zhang:2024bwf}, the HFRS, and a high-precision experimental spectrometer ring (SRing) \cite{Yan:2023qeu}.
The 114-meter-long iLinac is equipped with three ion sources, including a new-generation 45 GHz, 20 kW superconducting electron cyclotron resonance(ECR) ion source. 
It can accelerate the $^{238}\mathrm{U}^{35+}$ beam to a maximum energy of 17 $\mathrm{MeV/u}$. The BRing has a circumference of 569 meters and a maximum magnetic rigidity of 34 Tm. 
Typical parameters of the beams from BRing are shown in Table \ref{tab:beam_param}. 

The $^{238}\rm{U}^{35+}$ ions can be accelerated in the BRing to 835 MeV/u, with an intensity of about $1.0\times 10^{11}$ ppp. 
These high-energy and high-intensity beams can be extracted from the BRing via either slow or fast extraction modes and sent to different experimental terminals, for investigations in various fields such as nuclear physics, atomic physics, and muon tomography. 
In the fast extraction mode, the beam has a repetition rate of 3 Hz. 
In a typical slow extraction mode, the beam has an extraction cycle of 2 s and an extraction time of 1 s. 
 
\begin{figure*}[!htb]
	\centering
	\includegraphics
	[width=1.0\textwidth]{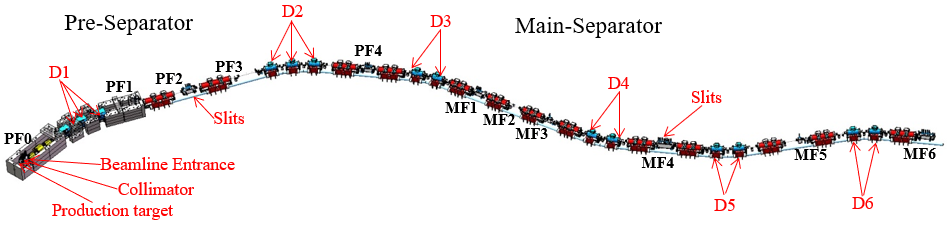}
	\caption{Schematic layout of the HFRS. HFRS is divided into two stages including the Pre-Separator and Main-Separator. The Pre-Separator is from $\mathrm{PF}0$ to $\mathrm{PF}4$, and the Main-Separator is from $\mathrm{PF}4$ to $\mathrm{MF}6$. The blue components are the dipole magnets, and the red components are the quadrupole magnets.}
	\label{fig:HFRS}
\end{figure*}

As an important experimental terminal at HIAF, HFRS is an in-flight separator at relativistic energies and is designed for the investigation of rare isotopes far away from the line of beta stability\cite{Sheng:2023ojn}. 
The rare isotope beams are produced, collected, purified and transported by HFRS beamlines.
As shown in Fig. \ref{fig:HFRS}, a two-stage structure is designed for HFRS. The first stage is from $\mathrm{PF}0$ to $\mathrm{PF}4$, which is generally denoted as Pre-Separator. 
In Pre-Separator, $\mathrm{PF}0$ is the cave where the target is located and bombarded by the primary beam extracted from BRing.
The Pre-Separator is $74$ meters long and can be used to eliminate the primary beams as well as undesired fragments. 
The second stage is from $\mathrm{PF}4$ to $\mathrm{MF}6$, which is generally called Main-Separator.
The Main-Separator can be used for the further purification of the nuclides of interest and the particle identification of reaction products.

The representative parameters of the HFRS beamlines are given in Table \ref{tab:hfrs_param}.  
The momentum acceptance of HFRS can be adjusted by using the slits at PF2 and MF4.
The maximum magnetic rigidity of HFRS is 25 $\mathrm{Tm}$, which enables it to transport pions/muons with momenta up to 7.5 $\mathrm{GeV/c}$.  
The total length of HFRS is 192 $\mathrm{m}$, approximately equal to the decay length of pion with a momentum of 3.4 $\mathrm{GeV/c}$.
Further details of the HFRS can be found in Ref.\cite{Sheng:2023ojn}. 

\begin{table}[h]
	\centering
	\caption{Main parameters of the HFRS.}
	\label{tab:hfrs_param}
	\begin{tabular}{lc}
		\hline
		\hline
		Parameters &  Value \\
		\hline
		Length [m] &  192 \\
		RMS of beam spot at the target [mm]  & $\sigma_x=0.4$\;,\;$\sigma_y=0.6$  \\
		Max. magnetic rigidity [Tm] &  25 \\
		Max. angular acceptance [mrad] &  x'=$\pm$30\;,\;y'=$\pm$25 \\
		Max. momentum acceptance  &   $\pm$2\% \\
		\hline
		\hline
	\end{tabular}
\end{table}

\section{Production of muon beam}

In this section, we discuss the strategy to produce a muon beam based on HFRS. Firstly, the pion production with different projectiles is analysed. Then, the muon beam that could be obtained with HFRS is investigated and the suitable primary beams of HIAF for  producing muon beams at specific energies in the GeV range are explored.

\subsection{Pion production with different projectiles at HIAF}

In this subsection, the behavior of pion production with several representative bombarding beams, including protons, $^{18}\rm{O}^{6+}$, $^{78}\rm{Kr}^{19+}$, and $^{238}\rm{U}^{35+}$ will be discussed. 
The software G4Beamline\cite{G4B} is utilized for simulations, 
and the physics list QGSP-BERT is selected for modeling the interactions.
The parameters of the bombarding beams are adopted as shown in Table \ref{tab:beam_param}. 
For the bombarding beam, the beam spot follows a Gaussian distribution, with the standard deviations of 0.4 mm and 0.6 mm in the horizontal and vertical directions, respectively.

Regarding the target in $\mathrm{PF}0$, a cylindrical graphite target with a radius of 50 mm and a thickness of 100 mm is used. For different projectiles, the target dimensions remain fixed because of the space constraints of the $\mathrm{PF}0$ cave. The pions are recorded by a virtual detector placed 10 mm after the production target.

\begin{figure*}[!ht]
	\centering
	\includegraphics[width=1.0\textwidth]{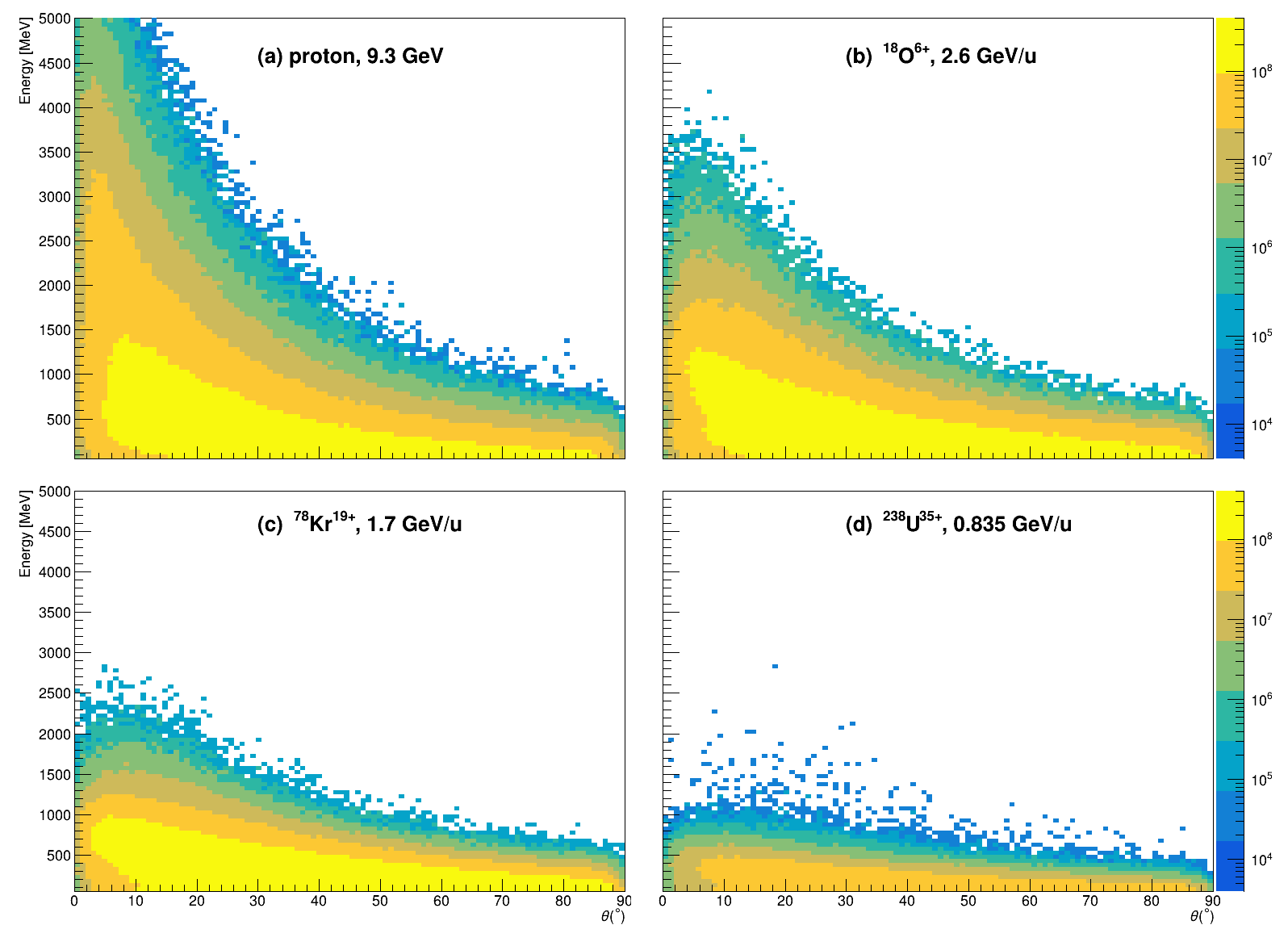}
	\caption{Kinetic energy vs scatter angle distribution of $\pi^+$ generated after the production target with different bombardimg beams as shown in Table \ref{tab:beam_param}. (a) pion yield $Y_{\pi^{+}}$ with proton beam; (b) pion yield $Y_{\pi^{+}}$ with $^{18}\rm{O}^{6+}$ beam; (c) pion yield $Y_{\pi^{+}}$ with $^{78}\rm{Kr}^{19+}$ beam; (d) pion yield $Y_{\pi^{+}}$ with $^{238}\rm{U}^{35+}$ beam.}
	\label{fig:2}
\end{figure*}

The energy and emission angle distributions of the $\pi^+$ generated by the beams specified in Table \ref{tab:beam_param} and subsequently recorded by the virtual detector located downstream of the target are presented in Fig. \ref{fig:2}, while these of the $\pi^-$ show similar characteristics.
It is evident that the products exhibit a pronounced forward recoil phenomenon. Consequently, a zero-degree spectrometer can be used to collect these products. 
From Fig. \ref{fig:2}, it is clear that there is an effect on the pion yield $Y_{\pi}$ for $\pi^+$ and $\pi^-$ due to the different energy and flux intensities of the primary beam available at HIAF, as shown in Table \ref{tab:beam_param}. 
Considering the parameters of the HIAF projectile beams, the pions generated from the primary proton beam exhibit a significantly higher maximum energy, making them more suitable for producing higher energy muon beams. For instance, if a high-energy muon beam with $E_\mu$ exceeding 3 GeV is required, the proton beam stands out as the best option for HIAF. On the other hand, while the pions generated from ion beams have a lower maximum energy, they have higher intensities in certain phase spaces, which makes them a better choice for producing muon beams in these specific regions. 

\begin{figure}[h]
	\centering
	\subfigure{
	\label{fig:3a}
	\includegraphics[width=0.45\textwidth]{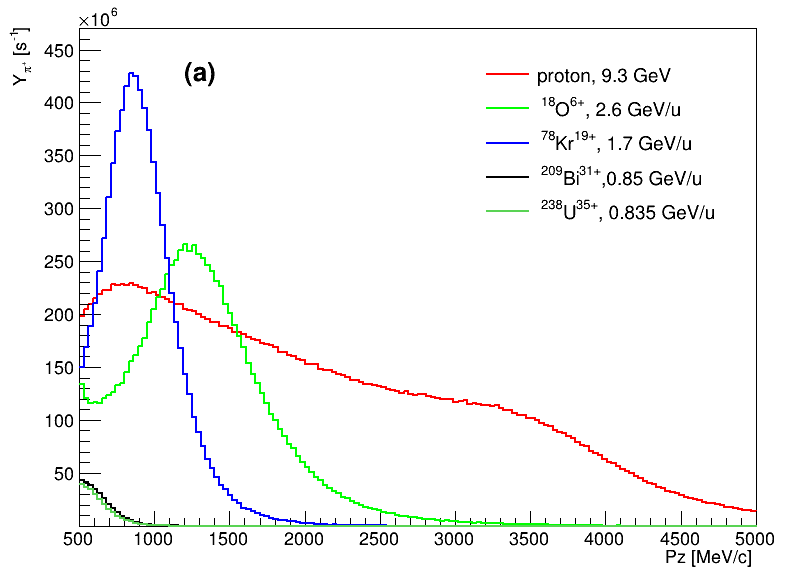}
	}
	\subfigure{
	\label{fig:3b}
	\includegraphics[width=0.45\textwidth]{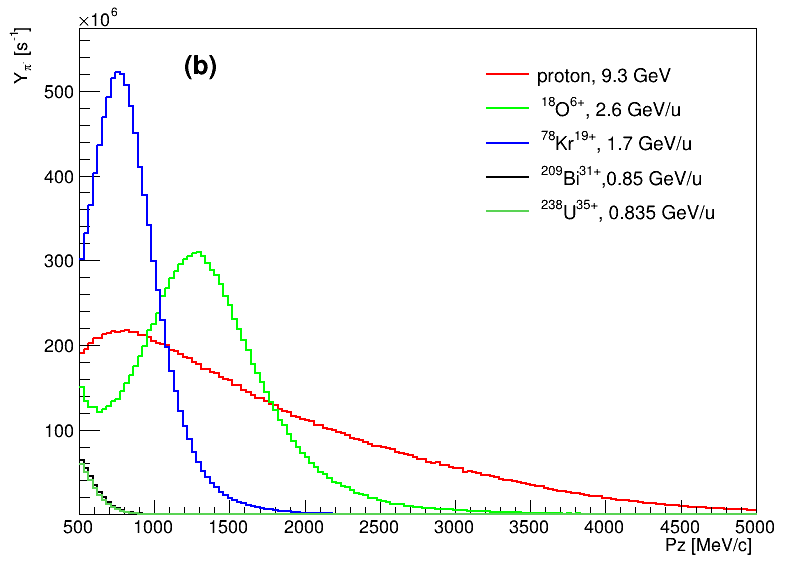}
	}
	\caption{Pion yield recorded by the virtual detector positioned at the beamline entrance with different ions. (a) Pion yield distribution of $\pi^+$; (b) Pion yield distribution of $\pi^-$.}
	\label{fig:3}
\end{figure}

Due to the large phase space of pion generation and limited angular and momentum acceptance of the HFRS beamline, only a small fraction of the pions can enter the beamline. 
Fig. \ref{fig:3} shows the momentum spectra of the pions within the angular acceptance of HFRS beamline, clearly demonstrating the choice and suitability of different primary beams for producing pions at specific energies at HIAF. In the GeV energy range, the momentum of the muons from pion decay in flight is strongly correlated with that of the pions \cite{ParticleDataGroup:2024cfk}. Therefore, the suitable primary beams provided by HIAF for producing muon beams at specific GeV-range energy can be inferred from Fig. \ref{fig:3}.
Generally, for muon flux below 1.1 GeV/c, the best choice is $^{78}\rm{Kr}^{19+}$ beam. For momentum between 1.1 GeV/c and 1.6 GeV/c, the $^{18}\rm{O}^{6+}$ projectile is more suitable. For momentum higher than 1.6 GeV/c, the proton beam is the best option. Moreover, the pion yields with projectiles such as $^{209}\rm{Bi}^{31+}$ and $^{238}\rm{U}^{35+}$ are significantly lower than those of the other considered projectiles.
This conclusion applies to both $\pi^+$ and $\pi^-$.
In addition, it is noticed that the production rates for $\pi^+$ and $\pi^-$ are different, as shown in the plots. 
For the ion beams, the $\pi^-$ yield is slightly higher than the $\pi^+$ yield, whereas for the proton beam, the $\pi^-$ yield is lower than the $\pi^+$ yield. 
This difference is attributed to the fact that the $\pi^+$ production cross-section is higher than that of $\pi^-$ in proton collisions \cite{Chemakin:2001qx, E910:2007puw}, and lower than that of $\pi^-$  in heavy-ion collisions \cite{Norbury:2006hp, Werneth:2013iba}.

The above results indicate that at HIAF the projectiles of $^{238}\rm{U}^{35+}$ and $^{209}\rm{Bi}^{31+}$ are not suitable for producing GeV-energy muons. 
In the following analysis, we will focus the discussion on projectiles including $^{78}\rm{Kr}^{19+}$, $^{18}\rm{O}^{6+}$, and proton beams.

\subsection{The produced muon beams with different projectiles at HIAF}

The pions within the angular acceptance of HFRS beamline will decay to muons with a branching ratio of 99.99\% \cite{ParticleDataGroup:2024cfk}:

\begin{align}
\pi^+ \rightarrow \mu^+ + \nu_\mu \\
\pi^- \rightarrow \mu^- + \bar{\nu}_\mu
\end{align}

By setting up the magnetic rigidity of the HFRS, we can determine the charge and momentum of the $\pi/\mu$ beam, which is critical for meeting the requirements for different physics and application purposes. 
In this part of the analysis, the magnetic rigidity configuration of the entire HFRS is set to the same value according to the required muon beam momentum.
The muon beam profile and the obtained muon flux will be first discussed in detail, followed by the discussion on the energy spectrum of the generated muon flux.

\subsubsection{muon beam profile}

\begin{figure*}[!htb]
	\centering
	\includegraphics[width=1.0\textwidth ]{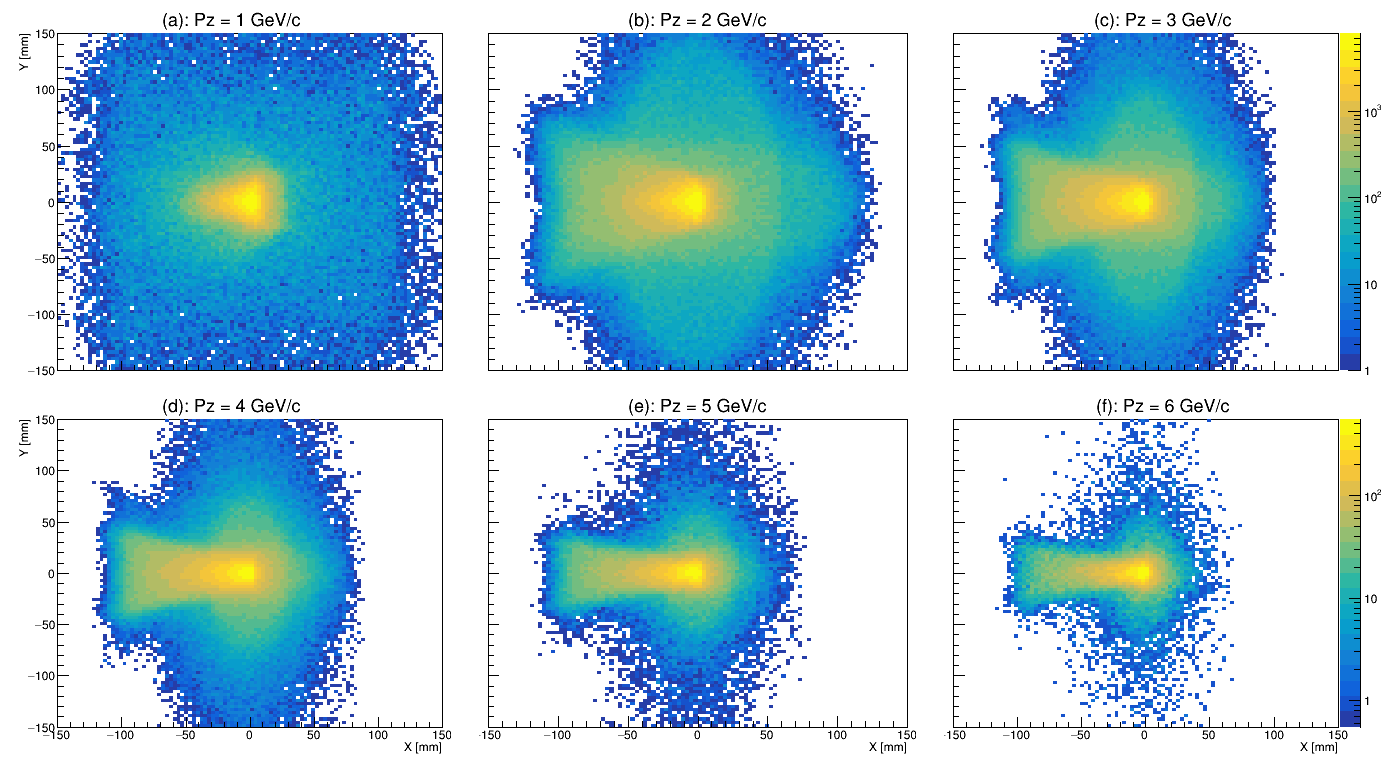}
	\caption{The muon beam profile with the momentum ranging from 1 to 6 GeV/c with a step size of 1 GeV/c at the exit of the HFRS. The projectile of proton at HIAF is adopted here.}
	\label{fig:muon_profile}
\end{figure*}

The size of the beam profile is a key parameter of the muon beam, especially for applications such as muon tomography.
In this subsection, the beam spots of the muon flux at several typical momenta at the exit of the HFRS beamline are investigated.
The results of $\mu^+$ are presented here, and those of $\mu^-$ are similar. 
The proton projectile is taken as an example here.

Generally, the size of the muon beam profile can be adjusted according to the physical requirements. In Fig. \ref{fig:muon_profile}, the representative muon beam profiles with momenta ranging from 1 to 6 GeV/c with a step size of 1 GeV/c are shown respectively.
Owing to the impact of the high-order effects of the HFRS beam optics, the muon beam spot shows a wing-shaped distribution. In addition, as the momentum of the muon increases, the distribution of the beam spot gradually contracts. This is mainly because the HFRS beam optics is optimized at the maximum magnetic rigidity. Compared with the case of the maximum magnetic rigidity, the magnets of the HFRS differ in their effective lengths and edge-field distributions under lower magnetic rigidity conditions, and these differences become increasingly prominent as the magnetic rigidity value deviates further from its maximum. This leads to a degradation of the focusing effect of the beam spot at the exit of the HFRS. In the future, we will optimize the beam optics for various magnetic rigidity values to achieve an excellent focusing effect. Also, the distribution shape and size of the muon beam spot at the exit of the HFRS can be adjusted according to future physical requirements.

\begin{table}[h]
	\centering
	\caption{Beam spot size as indicated by the ratio $R_{\mu}(x)$ is shown for several muon beam energies. 
$R_{\mu}(x)$ is defined as the number of muons in the central squared region with side length of $x$(mm) divided by the total number of muons. The number of muons is obtained at the exit of the HFRS. The 9.3 GeV proton primary beam was used in these simulations.
}
	\label{tab:ratio_muon}
	\begin{tabular}{lccccc}
		\hline
		\hline
		$P_{\mu}$ & 20 mm & 40 mm & 60 mm & 80 mm & 100 mm \\
		\hline
		1 GeV/c & 63.1\% & 79.0\% & 82.9\% & 85.2\% & 86.8\% \\
		2 GeV/c & 72.2\% & 85.4\% & 88.0\% & 89.9\% & 91.7\% \\
		3 GeV/c & 76.9\% & 89.4\% & 91.5\% & 93.2\% & 94.5\% \\
		4 GeV/c & 80.5\% & 91.9\% & 93.6\% & 94.9\% & 95.9\% \\ 
		5 GeV/c & 82.4\% & 93.6\% & 95.0\% & 96.0\% & 96.7\% \\
		6 GeV/c & 83.8\% & 94.8\% & 95.9\% & 96.7\% & 97.3\%\\
		\hline
		\hline
	\end{tabular}
\end{table}

Table \ref{tab:ratio_muon} shows the ratio $R_{\mu}(x)$, which is defined as the number of muons in the central square region with a side length of $x$ (mm) divided by the total number of muons at the exit of the HFRS.
In this study, we discuss the size of the beam profile when $R_{\mu}(x)$  exceed 90\%, which is suitable for muon tomography.
From Table \ref{tab:ratio_muon}, we find that for low energy muon beam generated mainly by the heavy-ion projectiles, it requires a side length of around 100 mm to reach 90\% collection efficiency; In contrast, for high energy muon beam, a side length of less than 40 mm is enough for 90\% collection efficiency. Moreover, as the side length of the central square region increases, the corresponding ratio increases slightly.

\subsubsection{muon yield}

In this subsection, the intensities of the muon flux as well as the corresponding purities at different momenta from the exit of the HFRS will be presented.
Both cases of $\mu^+$ and $\mu^-$ flux are discussed.

\begin{figure}[!htb]
	\centering
	\subfigure{
	\label{fig:5a}
	\includegraphics[width=0.5\textwidth]{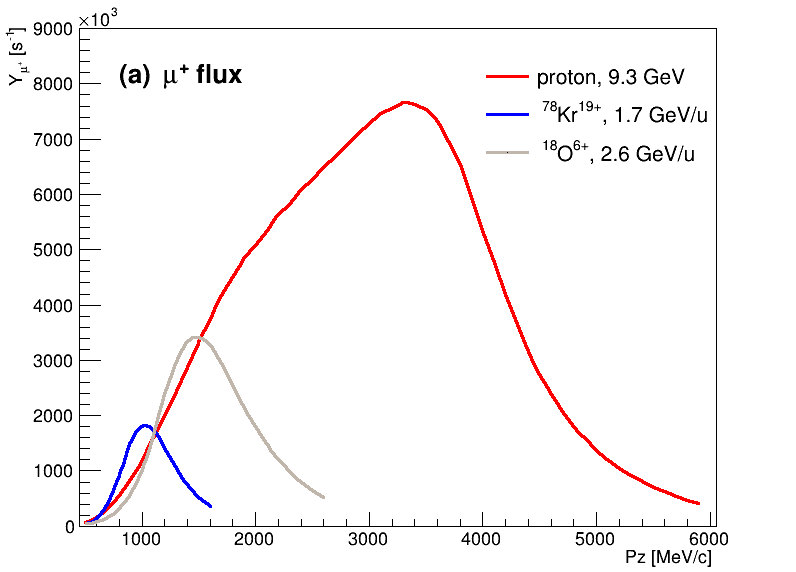}
	}
	\subfigure{
	\label{fig:5b}
	\includegraphics[width=0.5\textwidth]{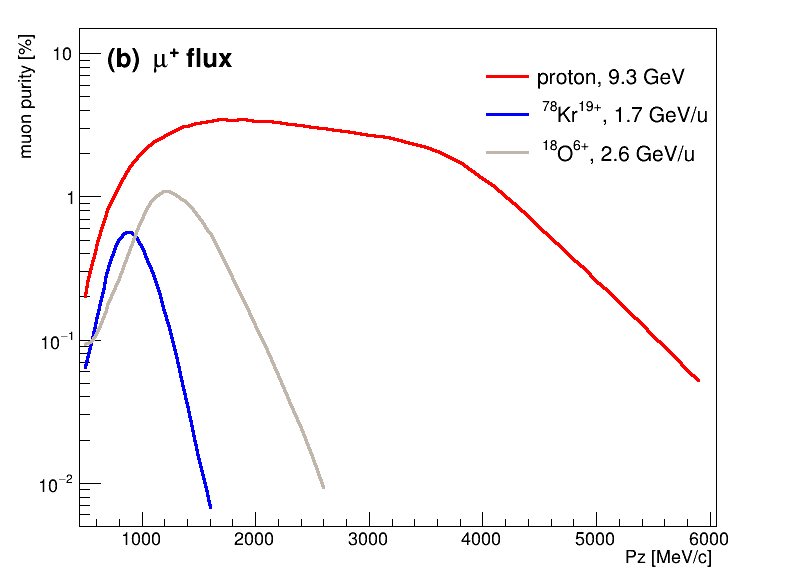}
	}
	\caption{(a) The intensity of the $\mu^+$ flux as a function of the muon momentum. (b) The muon purity of the $\mu^+$ flux as a function of the muon momentum. 
	 The $\mu^+$ flux is obtained at the exit of the HFRS. Results with different projectiles are given.  }
	\label{fig:muonp_flux_ions}
\end{figure}

\begin{table}[h]
	\centering
	\caption{The maximum muon flux intensities with proton, $^{18}\rm{O}^{6+}$, and $^{78}\rm{Kr}^{19+}$ projectiles, and the corresponding muon beam momenta and purities.}
	\label{tab:param_muon_flux}
	\begin{tabular}{lccc}
		\hline
		\hline
		 & proton & $^{18}\rm{O}^{6+}$ & $^{78}\rm{Kr}^{19+}$ \\
		\hline
		$\mu^+$ beam \\
		\hline
		momentum [GeV/c] & $3.5~$ & $1.5~$ & $1.0~$ \\
		flux intensity [$\mu^+/s$] & $8.2\times 10^{6}$ & $3.5\times 10^{6}$ & $1.8\times 10^{6}$  \\
		muon purity  & 2.0\% & 0.80\% & 0.60\%  \\
		\hline
		$\mu^-$ beam \\
		\hline
		momentum [GeV/c] & $2.3~$ & $1.5~$ & $1.0~$ \\
		flux intensity [$\mu^-/s$] & $3.8\times 10^{6}$ & $4.2\times 10^{6}$ & $1.6\times 10^{6}$  \\
		muon purity  & 13\% & 20\% & 23\%  \\ 
		\hline
		\hline
	\end{tabular}
\end{table}

\begin{figure}[!htb]
	\centering
	\subfigure{
	\label{fig:6a}
	\includegraphics[width=0.5\textwidth]{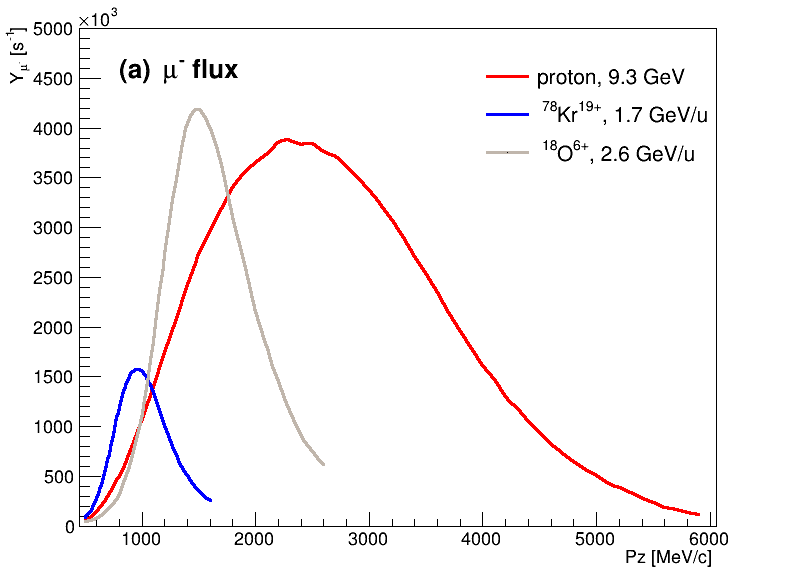}
	}
	\subfigure{
	\label{fig:6b}
	\includegraphics[width=0.5\textwidth]{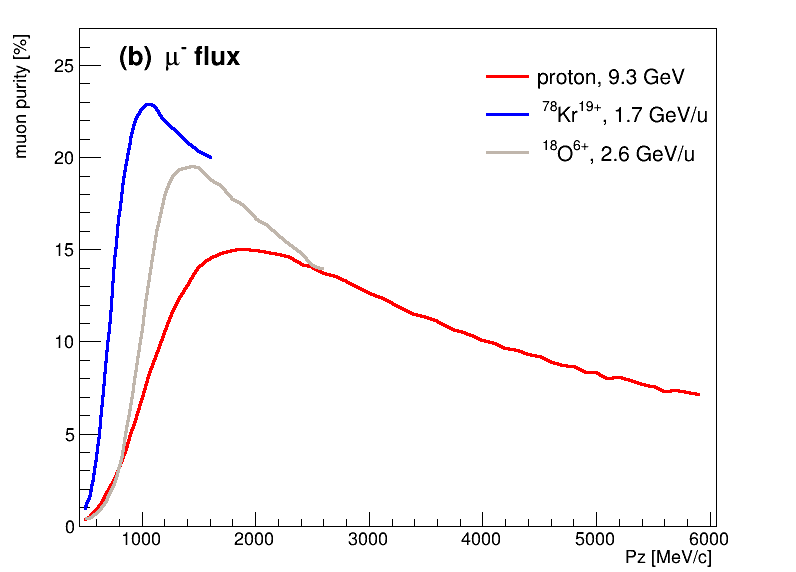}
	}
	\caption{(a) The intensity of the $\mu^-$ flux as a function of the muon momentum. 
	(b) The muon purity of the $\mu^-$ flux as a function of the muon momentum. 
	The $\mu^-$ flux is obtained at the exit of the HFRS. Results with different projectiles are given. }
	\label{fig:muonm_flux_ions}
\end{figure}

In Fig. \ref{fig:muonp_flux_ions} and Fig. \ref{fig:muonm_flux_ions}, the intensity as well as the purity of the muon flux is shown as a function of the muon momentum. The muon flux is obtained at the exit of the HFRS, and the purity is defined as the number of muons divided by that of all charged particles. Results for different projectiles are presented.
The maximum muon flux intensities with the corresponding muon beam energies and purities are listed in Table \ref{tab:param_muon_flux}.
For the $\mu^-$ flux at HFRS, the purity is significantly higher than that of the $\mu^+$ flux, and the reason will be explained in the following section.

These results indicate that, with the magnetic rigidity of the entire HFRS beamline set to the same value according to the desired muon beam momentum $p$ with the formula $B\rho[\mathrm{Tm}] = p[\mathrm{MeV}/\mathrm{c}]/300\,q$, the intensities of both the $\mu^+$ and $\mu^-$ fluxes can reach the order of $10^{6}$ $\mu$/s, given the projectile parameters provided by HIAF. However, the purity of the $\mu^{-}$ is around $20\%$, and the purity of the $\mu^{+}$ is at only a few percent.

\section{\label{sec:purification}purfication of the muon beam}

When a beam particle bombards the graphite target, different kinds of charged particles, including electrons, protons, mesons and so on, will be produced. 
Due to the overlapping of their magnetic rigidity values, these charged particles will mix with the muon beam and be transported forward along the HFRS beamline. 
As a result, the composition of the beam obtained at the exit of HFRS is complicated and the purity of $\mu^+$ and $\mu^-$ are constrained, as presented in the previous section. However, in many physics and application experiments, there is a quite high requirement on muon beam purity. Consequently, it is necessary to purify the obtained muon beam.
In the following discussions, the types of background particles in the muon beam with a momentum of $1.5~$GeV/c, where the $^{18}\rm{O}^{6+}$ projectile has its maximum muon flux intensity, and its purification scheme will be discussed as an example.

\subsection{Composition of the beam obtained at the exit of HFRS}

\begin{figure}[!htb]
	\centering
	\includegraphics[width=0.5\textwidth ]{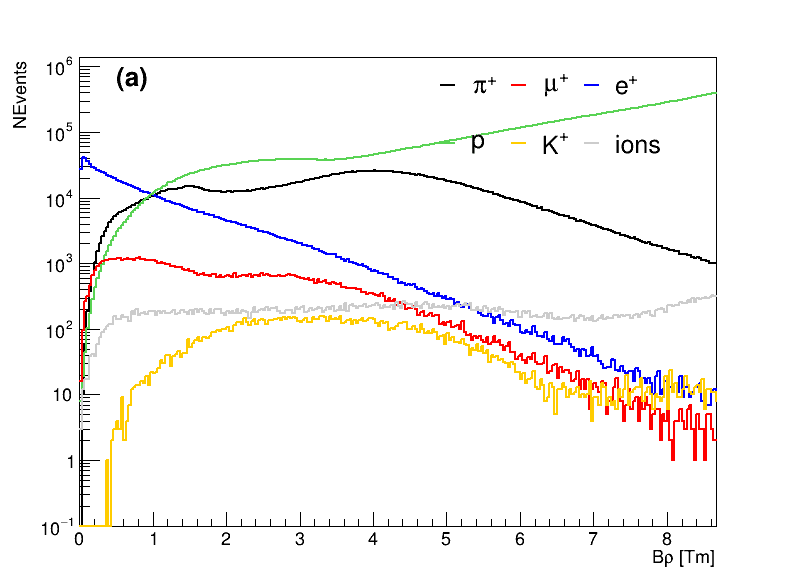}
	\includegraphics[width=0.5\textwidth ]{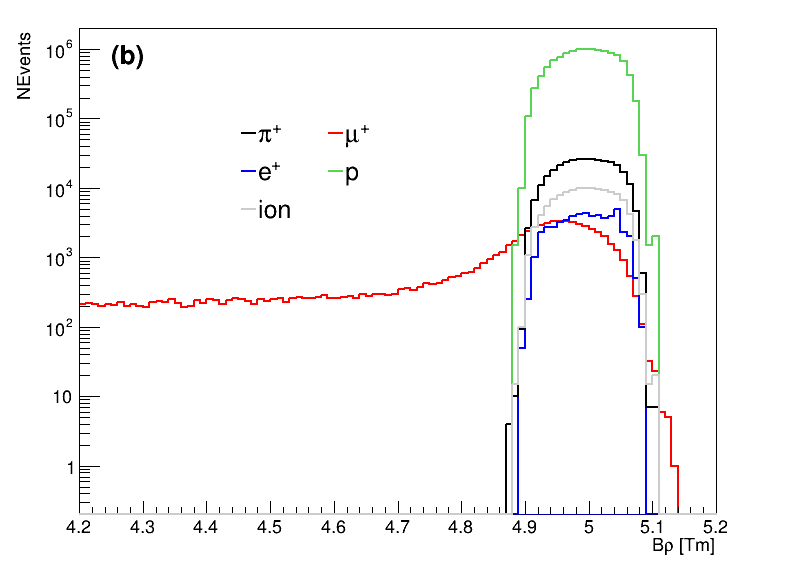}
	\caption{(a) The $\mu^+$ beam compositions at the entrance of the HFRS beamline.  (b) The $\mu^+$ beam compositions at the exit of the HFRS beamline. The projectile of 2.6 GeV/u $^{18}\rm{O}^{6+}$ beam is used here. The magnetic rigidity value of the whole HFRS is set to 5 Tm, corresponding to muon momentum of $1.5~ \mathrm{GeV/c}$. }
	\label{fig:muon_comp}
\end{figure}

As shown in the upper panel of Fig. \ref{fig:muon_comp}, in the plotted magnetic rigidity region, the background particles for the $\mu^+$ beam at the entrance of HFRS consist of $\pi^{+}$, $e^+$, protons, $K^+$ and heavy ions with similar magnetic rigidities. Protons, $\pi^+$ and $e^+$ are the dominant background particles. 
The production rate of $e^+$ decreases rapidly as the momentum increases, while that of protons increases significantly.
A small fraction of the heavy-ion fragments produced during the target interaction may also share the same magnetic rigidity as muons. These fragments usually have intensities two to three orders of magnitude lower than those of the protons.
The muons resulting from the in-flight decay of pions and kaons, as well as the  aforementioned beam components accepted by the HFRS, will be transported to the end of the beamline, provided their magnetic rigidities align with the beamline configuration.
Consequently, poor purity of the muon beam at the exit of HFRS can be anticipated, given that the magnetic rigidity parameter is set to the same value for the entire HFRS beamline.

In the lower panel of Fig. \ref{fig:muon_comp}, the composition of the $\mu^+$ beam obtained at the exit of HFRS is shown. Herein, the magnetic rigidity parameter is set to $5~ \mathrm{Tm}$, which corresponds to a muon momentum of $1.5 $ GeV/c. The uncertainties of the background particles' magnetic rigidity values are $\pm$ 2\%, which are determined by the design of the HFRS beamline. For the muon, there is an additional long tail extending toward the low magnetic rigidity region, mainly due to the in-flight decay of the pions.
The number of obtained protons is almost two orders of magnitude greater than that of the obtained $\mu^+$.
It is evident that the beam is almost full of protons.
In addition, the number of $\pi^+$ is approximately one order of magnitude higher than that of $\mu^+$. 
The number of $e^+$ is comparable to that of $\mu^+$ and is approximately two orders of magnitude lower than that of protons.
Therefore, only about $1\%$ of the particles are muons in the obtained beam, which is in accordance with the results shown in Fig. \ref{fig:muonp_flux_ions}.

\begin{figure}[!htb]
	\centering
	\includegraphics[width=0.5\textwidth ]{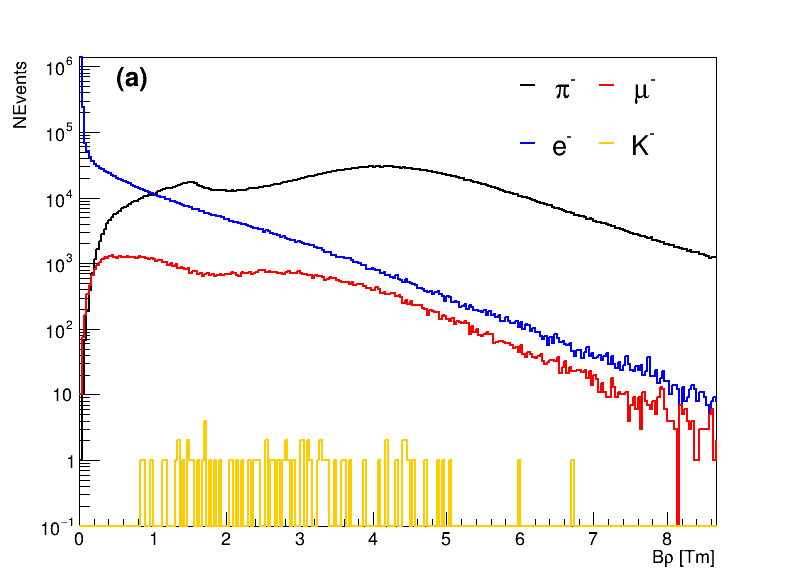}
	\includegraphics[width=0.5\textwidth ]{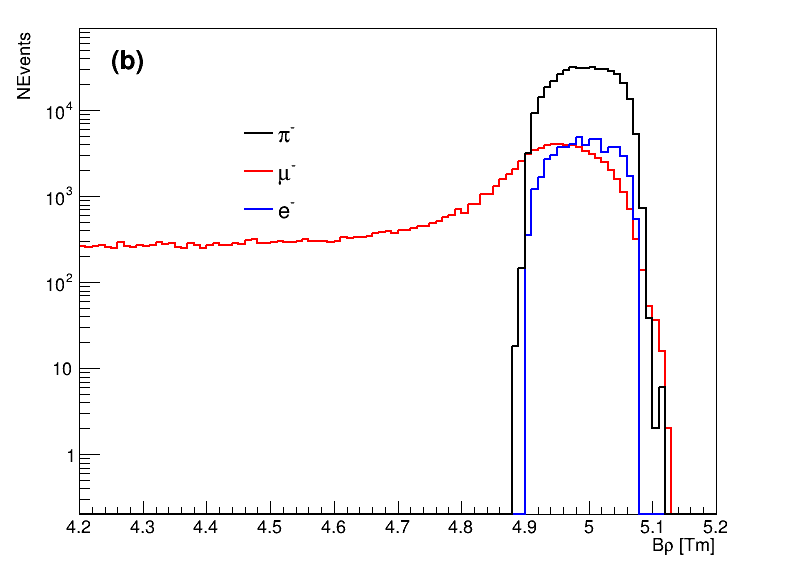}
	\caption{(a) The $\mu^-$ beam compositions at the entrance of the HFRS beamline. (b) The $\mu^-$ beam compositions at the exit of the HFRS beamline. The projectile of 2.6 GeV/u $^{18}\rm{O}^{6+}$ beam is used here. The magnetic rigidity of the entire HFRS is set to 5 Tm, corresponding to muon momentum of $1.5~ \mathrm{GeV/c}$. }
	\label{fig:muon_spec}
\end{figure}

For $\mu^-$, the dominant background particles consist mainly of $\pi^-$, $e^-$ and $K^-$, as can be seen from the beam compositions at the entrance of HFRS shown in the upper panel of Fig. \ref{fig:muon_spec}. Due to its opposite charge, the proton, which is the most dominant background particle for $\mu^+$, is not accepted by the HFRS beamline at all. Since the charge states of the produced heavy ions are basically positive, heavy ions are also absent as background particles for the $\mu^-$ beams. 
On the other hand, the intensities of $K^-$ are several orders of magnitude lower than those of $\mu^-$, so their contribution can be neglected.
The production rates of $\pi^-$ and $e^-$ are similar to those of $\pi^+$ and $e^+$.
Since the number of background particles for the $\mu^-$ beams is much smaller than that for the $\mu^+$ beams, the purity of the $\mu^-$ beams is expected to be higher than that of the $\mu^+$ beams, which can be inferred from the spectrum of the beam compositions at the exit of the HFRS beamline shown in the lower panel of Fig. \ref{fig:muon_spec}.
The beam purity for the $\mu^-$ is about $20\%$, as can be seen from Fig. \ref{fig:muonm_flux_ions}.

The above results indicate that, when the magnetic rigidity parameters of the entire HFRS are set to the same value, the background particles with suitable momenta, along with muons from in-flight decay of pions, are transported together through the HFRS beamline. This significantly reduces the muon beam purity. 
In particular, the presence of background protons seriously exacerbates this issue for  $\mu^+$, making the corresponding beam purity for $\mu^+$ as low as about $1\%$, which is unsatisfactory for applications such as muon tomography.

\subsection{Purification scheme for the muon beam}

The HFRS selects the beam particles by setting the magnetic rigidity parameter, which is proportional to the particles' momenta. As discussed above, for the muon beam generated from the decay of pions, the undecayed pions are always one of the main contaminants in the beam. 
In order to suppress the background particles, we are studying a scheme based on the momentum differences between muons and other charged particles. 

\begin{figure}[!htb]
\centering
\includegraphics
  [width=0.5\textwidth ]{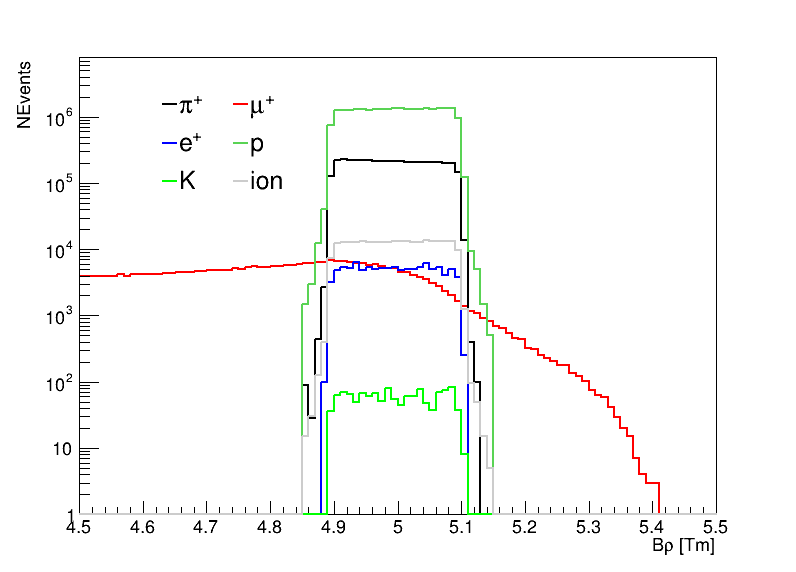}
\caption{The composition of the $\mu^+$ beam at the PF2 point of the HFRS beamline, with the magnetic rigidity value of the first section of HFRS ($B\rho_1$) set to $5.0~$Tm (the corresponding $\pi/\mu$ momentum of $1.5~$GeV/c).}
\label{fig:muon_profile_D1}
\end{figure}

Taking the $\mu^+$ beam with a momentum of 1.5 GeV/c as an example, the spectrum of various particles recorded at the focal point PF2 downstream of the first dipole magnet is presented in Fig.  \ref{fig:muon_profile_D1}. The charged particles such as pions, kaons, electrons, and protons are restricted within a range of approximately $\pm 2\%$ of 5 Tm, while the muon spectrum has a long low-energy tail and a high-energy tip. The muons outside the momentum selection region mainly come from the decay of pions inside the first dipole magnet D1 and along the beamline from D1 to PF2. This observation reminds us that the muon beam can be purified according to the differences between the muons and other background particles. This is also the strategy used in the nuSTORM project to obtain a pure muon beam\cite{Liu:2015opa}.
Therefore, we develop the strategy by dividing the HFRS into two sections.
The magnetic rigidity of the second section of the HFRS is set according to the requirement of the muon beam. For instance, if we need a muon beam with a momentum of 1.5 GeV/c, we can set the magnetic rigidity value to 5 Tm.
In contrast, the magnetic rigidity of the first section has a slightly different value. Details of this strategy is discussed as follows.

\begin{figure}[!htb]
\centering
\includegraphics
  [width=0.5\textwidth]{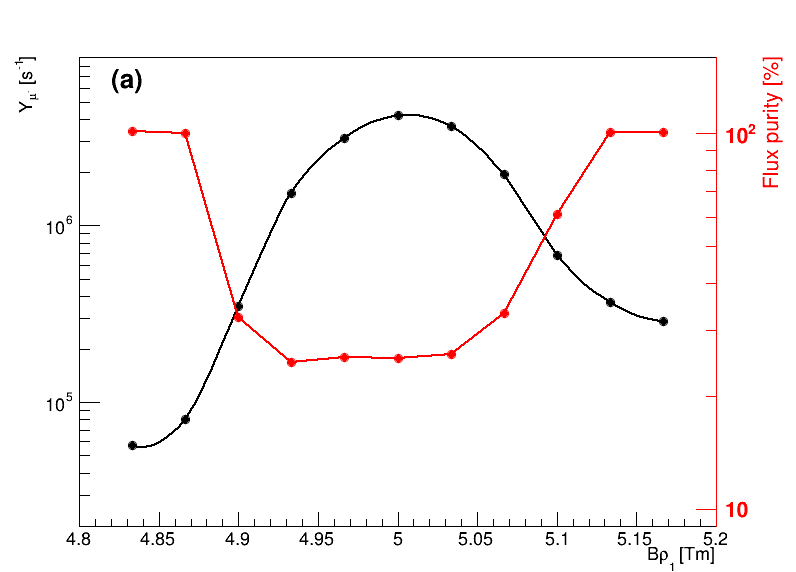}
\includegraphics
  [width=0.5\textwidth]{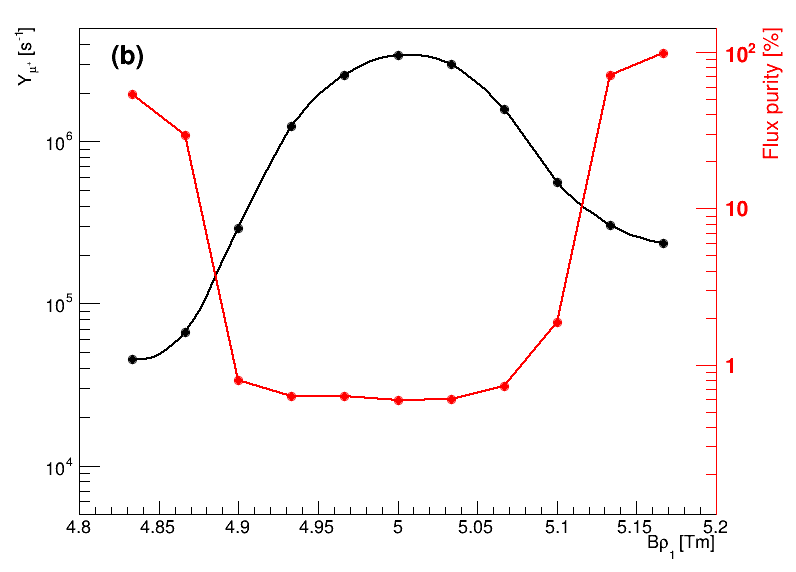}
\caption{The obtained muon intensity and purity by scanning the magnetic rigidity value of the first section of the HFRS components ($B\rho_1$). The magnetic rigidity value of the section of the HFRS components ($B\rho_2$) is set at 5 Tm.
(a) The performance for $\mu^-$ flux; (b) The performance for $\mu^+$ flux.}
\label{fig:tune}
\end{figure}

In order to obtain a pure muon beam with a momentum of 1.5 GeV/c, we set the magnetic rigidity value of the second section of HFRS, which is denoted as $B\rho_2$, to 5 Tm. 
Then, we scan the magnetic rigidity value of the first part of HFRS, which is denoted as $B\rho_1$, to obtain the optimal value with respect to both muon intensity and muon purity. The boundary point of the two sections is set at the focal point PF2.
The scanning results are shown in Fig. \ref{fig:tune}.  
It can be observed that when the values of $B\rho_1$ and $B\rho_2$ are identical, both are 5 Tm,  the purity of the beam is the lowest while the muon yield reaches its highest value. 
With the gradual increase of $B\rho_1$, the purity of the muon beam rises gradually, while its yield drops steadily. 
For the $\mu^-$ beam, when the difference $\Delta B\rho = B\rho_1 - B\rho_2$ reaches 0.13 Tm, the beam purity reaches 100\% with a corresponding yield of $3.7\times 10^5 ~ \mu^-$/s. For the $\mu^+$ beam, $\Delta B\rho$ needs to reach 0.17 Tm to reach $100\%$ purity, and the corresponding yield is $2.4\times 10^5 ~ \mu^-$/s. 
The $\mu^+$ beam requires a larger difference because the number of protons in the beam is more than an order of magnitude higher than that of other charged particles (as shown in Fig. \ref{fig:muon_profile_D1}), so a larger value is required to eliminate the protons in the beam. Furthermore, we can also reduce the value of $B\rho_1$, but the muon yield decreases much faster when the purity of the muon beam gradually increases. This asymmetry of the muon yield and purity results from the asymmetry of the spectrum of the muon, which is shown in Fig. \ref{fig:muon_profile_D1}.

In Fig. \ref{fig:tune}, as a purification example, the boundary point for different magnetic rigidity settings is set at the focal point PF2. Certainly, the boundary point can be alternately set at other focal points, such as PF4, MF1, MF2, and so on. The greater the distance between the boundary point and the primary target, the longer the decay length of the pions, and the more muons are produced. However, the loss rate of muons also increases due to the difference in magnetic rigidity distribution between the decaying muons and pions. Therefore, for a muon beam with a specific momentum, the determination of the optimal location for the boundary point to obtain a high-intensity and high-purity beam is a complex task. It will be carried out through simulations that take into account the specific physical requirements. Notably, to obtain a muon beam with controllable momentum spread, the boundary point must be set within the pre-separator of the HFRS. In this way, the muons generated by the in-flight decay will pass through the entire point-to-point optical system of the main-separator in subsequent transmission. At the dispersion plane MF4, the momentum of these muons will have a linear correlation with their horizontal positions. By adjusting the width of the momentum slit at MF4, effective control of the momentum spread of the muon beam can be achieved. Conversely, if the boundary point is not set in the pre-separator, the momentum spread of the muon beam will become uncontrollable and will present a wider distribution at the exit of the HFRS,  as shown in Fig. \ref{fig:muon_profile_D1}.

\section{Particle Identification}

The beam purification strategy in Sec. \ref{sec:purification} can effectively purify the obtained muon beam, but it leads to a significant loss in muon intensity, which is undesirable, especially when a large amount of statistics is needed. 
Alternatively, if muons can be distinguished from other charged particles such as protons, pions, and electrons, this approach can also meet the needs of most experiments without sacrificing muon intensity.
In this section, we explore the feasibility of discriminating muons from these other charged particles.

\begin{figure}[!htb]
\centering
\includegraphics
  [width=0.45\textwidth ]{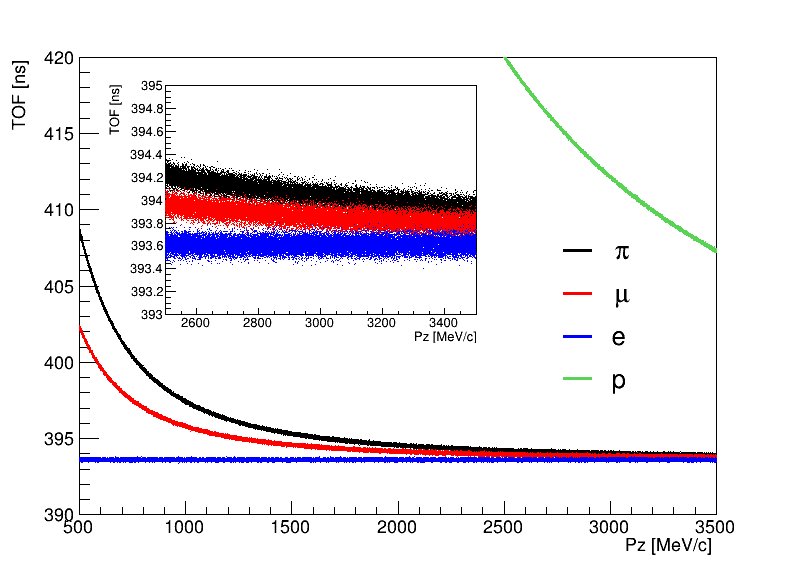}
\caption{The TOF distribution of the charged particles as a function of momentum. The two TOF detectors are located at PF4 and MF6.}
\label{fig:TOF1}
\end{figure}

Beam experiments require different particle identification methods at different energy ranges.
For low-energy cases, measuring the time-of-flight (TOF) is a common particle identification approach.
On the HFRS beamline, at points PF4 and MF6, several TOF detectors are installed to measure the TOF value of charged particles. The distance between these two points is  $118~$m.  
In Fig. \ref{fig:TOF1}, the TOF distribution of the charged particles as a function of momentum are shown.
We consider a momentum acceptance of $\pm 2\%$ and assume a TOF detector time resolution of 50 ps.
As expected, the TOF differences between particles decrease with increasing particle momentum.
Protons, having a significantly larger mass than other particles, have a clearly distinguishable TOF from other charged particles.
In the plotted momentum or magnetic rigidity range, the proton's TOF is at least 10 ns longer than that of other particles.
At low momentum, $\pi$, $\mu$ and $e$ are easily distinguishable as their TOF differences are several nanoseconds.
However, at higher momenta, e.g., $2.5~$GeV, the TOF difference between $\pi$ and $\mu$ is about $200~$ps, and that between $\mu$ and $e$ is about $400~$ps.   
As a result, some muons can't be distinguished from other particles, as shown in the inset of Fig. \ref{fig:TOF1}.

The investigations above suggest that, given the current HFRS parameters, placing two TOF detectors with time resolution of $50~$ps at PF4 and MF6 respectively can make the TOF method an effective approach for identifying muons from other charged particles at momenta lower than $2~$GeV. At higher momenta, a more competitive method is required. This issue will be further discussed in our future work.

\section{Conclusion}

In this paper, a muon source based on the HFRS beamline of HIAF is thoroughly discussed.
The pion production with different projectiles is first presented, suggesting the appropriate choices of projectiles at HIAF for GeV-energy muon production. 
Regarding the obtained muon beam, the results show that muon intensity can reach the order of $10^{6} ~ \mu$/s, with a purity of $1.0\%$ for $\mu^+$ flux and a purity of $20\%$ for $\mu^-$ flux. 
In order to enhance the purity of the muon beam, two strategies are introduced.
The results indicate that the purity of both $\mu^+$ and $\mu^-$ beams can be effectively improved. For the muon beam with a momentum of 1.5 GeV/c, the intensity will be of the order of $10^{5} ~ \mu$/s, with $100\%$ purity.

The HFRS muon beam provides significant opportunities for expanding the application areas of muon tomography. With an energy level in the GeV range, this muon beam exhibits excellent penetration capabilities, achieving a range of approximately 1 meter in iron. The beam flux reaches the order of $10^{5} \sim 10^6 ~ \mu$/s, highlighting its great potential for rapid non-destructive testing of large objects. Furthermore, the energy spread of the HFRS muon beam can be better than $2\%$, enabling superior performance in density resolution and high atomic number (Z-value) resolution compared to cosmic ray muons. This feature makes it highly valuable for the rapid detection of high-Z materials. 
Moreover, this GeV-energy muons are also expected to serve as a platform for investigating various new physics beyond the Standard Model. It also should be noted that, due to the low energy loss of muons in the GeV energy range, they are also suitable for detector calibration \cite{HIRSH2018122}.

In the next step, we will further explore the potential to achieve better muon beam with the HFRS beamline. On the one hand, the optical model of the HFRS beamline is optimized for the isotope beams, thus it is possible to get higher intensity muon beam with the new optical model optimized for the muon beam. In addition, we consider the beam differ in energy as well as ion identities and production differences, and we will separate how much of that dependence is on energy or ion type in the further studies. For the beam purification part, we will consider another common strategy, which is to include a material absorber (low-Z preferred) toward the end of the beam line. In this case, The absorber can stop pions and protons by nuclear interactions and reduce electrons through bremsstrahlung. The muons would have energy loss and scattering. And we will combine this strategy with the strategy described in this paper. 

Subsequently, an exploration of the potential for enhancing muon beams through the HFRS beamline will be conducted. Firstly, the optical model of the HFRS beamline, which was initially designed for isotope beams, will be re-optimized specifically for muon beams with the aim of increasing the beam intensity. 
Secondly, we will conduct a detailed and systematic investigation into the characteristics of muon beams, exploring the impact of the beam energy on a single projectile.
In terms of beam purification, an additional strategy will be employed. A low-Z material absorber will be installed at one of the focal points of the beamline, such as PF2 or MF4. This absorber will eliminate pions and protons through nuclear interactions and electrons via bremsstrahlung processes, although the muons would have energy loss and scattering. Furthermore, this strategy will be integrated with the methods described in this paper to achieve a more efficient purification process.

\section*{acknowledgments}		
This work was financially supported by High Intensity heavy-ion Accelerator Facility (HIAF) project approved by the National Development and Reform Commission of China, the National Natural Science Foundation of China (Grants No. 12105327, No. 12005221, No. 12405337, No. 12405402, No. 12325504, No. 12027809), the Guangdong Basic and Applied Basic Research Foundation, China (Grant No. 2023B1515120067), and State Key Laboratory of Nuclear Physics and Technology, Peking University (Grant No. NPT2025KFY07).
Computing resources were mainly provided by the supercomputing system in the Dongjiang Yuan Intelligent Computing Center.

\section*{DATA AVAILABILITY}	
The data that support the findings of this article are
openly available\cite{HIAF}. 


\bibliography{biblio}

\end{document}